\documentclass[12pt]{article}
\usepackage{graphicx}
\usepackage{booktabs}
\usepackage[table]{xcolor}
\usepackage{geometry}
\usepackage{amsmath}
\usepackage{multirow}
\usepackage{geometry}
\usepackage{doi}
\usepackage{authblk}
\usepackage{subcaption}
\usepackage{caption}
\usepackage{url}
\usepackage{float}  
\title{\textbf{One year of ASPEX-SWIS operation - Characteristic features, observations and science potential}}
\author[1]{Abhishek Kumar}
\author[1, 2]{Shivam Parashar}
\author[1]{Prashant Kumar}
\author[1]{Dibyendu Chakrabarty}
\author[1, 3]{Bhas Bapat}
\author[1]{Aveek Sarkar}
\author[1]{Manan S. Shah}
\author[1]{Hiteshkumar L. Adalja}
\author[1]{Arpit R. Patel}
\author[1]{Pranav R Adhyaru}
\author[1]{M. Shanmugam}
\author[1]{Swaroop B. Banerjee}
\author[1]{K.P. Subramaniam}
\author[1]{Tinkal Ladiya}
\author[1]{Jacob Sebastian}
\author[1]{Bijoy Dalal}
\author[1, 2]{Aakash Gupta}
\author[1]{M.B. Dadhania}
\author[1]{Santosh V. Vadawale}
\author[1]{Shiv Kumar Goyal}
\author[1]{Neeraj Kumar Tiwari}
\author[1]{Aaditya Sarda}
\author[1]{Sushil Kumar}
\author[1]{Nishant Singh}
\author[1]{Deepak Kumar Painkra}
\author[1]{Piyush Sharma}
\author[1]{Abhishek J. Verma}
\author[1]{P. Janardhan}
\author[1]{Anil Bhardwaj}

{\small
\affil[1]{Physical Research Laboratory, Ahmedabad-380009, India}
\affil[2]{Indian Institute of Technology, Gandhinagar-382355, India}
\affil[3]{Indian Institute of Science Education and Research, Pune-411008, India}
}
\date{}
\begin{document}
\maketitle
\begin{abstract}
The Aditya-L1 mission, India’s first dedicated solar observatory at the first Lagrange point of the Sun-Earth system, carries the Solar Wind Ion Spectrometer (SWIS) as part of the ASPEX payload suite. Even before settling down at the Halo orbit, SWIS has been delivering nearly continuous \textit{in-situ} measurements of solar wind ion spectra. Moments of the velocity distribution function (VDFs) have been calculated to derive the solar wind bulk parameters like density, bulk speed, temperature etc. Through this work, we evaluate the performance of SWIS(henceforth, AL1-ASPEX-SWIS) through comparisons with contemporaneous measurements from the Wind and DSCOVR missions. A detailed case study of the 07~August~2024 interplanetary coronal mass ejection (ICME) is presented where sharp transitions in bulk speed, thermal speed, and number density were well-aligned with independent observations, confirming the instrument’s capability in capturing dynamic solar wind features. Spectral analysis of kinetic fluctuations revealed a well-defined inertial range with a spectral slope consistent with magnetohydrodynamic (MHD) turbulence. Additionally, a 17-month statistical comparison (January~2024–May~2025) shows strong agreement in bulk velocity ($R^2 \approx 0.94$ with Wind), with expected variability in thermal speed and density due to inter-instrument differences. These results confirm the scientific utility of AL1-ASPEX-SWIS in monitoring both transient events and long-term solar wind conditions.
\end{abstract}

\section{Introduction}
The Indian Space Research Organisation (ISRO) created a significant milestone in heliophysics research with the launch of its first dedicated solar observatory, Aditya-L1\,\cite{tripathi2022aditya}, on 02~September~2023. After several earth-bound orbits, the satellite was injected into a trans-L1 trajectory on 19~September~2023. It was successfully inserted into a halo orbit around the Sun--Earth L1 Lagrange point on 6~January~2024. After the performance verification phase the duration of which varied from payload to payload, Aditya-L1\,(AL1) eventually started to provide nearly continuous, observations of the Solar corona, photosphere, chromosphere the \textit{in-situ} solar wind, energetic ions and magnetic field.

Aditya-L1 contributes to a broader global effort to understand the variability, origin, and structure of the solar wind---an ongoing challenge in heliophysics. The solar wind, a supersonic stream of plasma primarily composed of protons and electrons, is continuously emitted by the Sun into the heliosphere. Understanding its generation and dynamic behavior requires multi-point, high-resolution, and long-term observations.

Recent missions such as NASA's \textit{Parker Solar Probe} and ESA-NASA's \textit{Solar Orbiter} have probed the inner heliosphere closer to the Sun than ever before. In parallel, long-term solar wind monitoring has been continuing through earlier missions such as \textit{Wind} (launched in 1994) and \textit{DSCOVR} (Deep Space Climate Observatory, launched in 2015), both located near the L1 point. The \textit{Wind} spacecraft carries instruments like the 3DP-PLSP (Three-Dimensional Plasma Analyzer – Proton/Ion Electrostatic Analyzer - Low Spectra Proton) and SWE-FC (Solar Wind Experiment – Faraday Cup). These instrument (described in references \cite{ogilvie1995swe, lin1995wind}) have provided detailed measurements of ion distributions and bulk parameters. Similarly, \textit{DSCOVR}'s PlasMag-FC (Plasma Magnetometer - Faraday Cup)\,\cite{lotoaniu2022validation} instrument offers continuous monitoring of solar wind density, speed, and temperature at L1. Together, these missions have contributed critical data on the composition, and temporal variability of the solar wind across solar cycles, forming the foundation for space weather forecasting and early warning systems for geomagnetic storms\,\cite{wilson2021quarter}.

\begin{figure}[!t]
  \centering
  \includegraphics[width=0.75\textwidth]{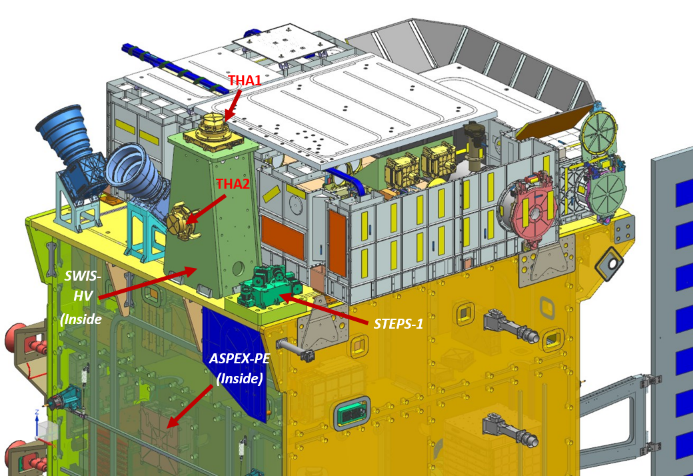}
  \caption{Configuration of payloads aboard Aditya-L1. The Solar Wind Ion Spectrometer (SWIS), a part of the ASPEX suite, is placed on the tower. Field-of-view (FOV) placement of the dual THA sensors (THA-1 and THA-2) aboard Aditya-L1. THA-1 and THA-2 are mounted orthogonally, allowing measurement of solar wind ion anisotropies and ensuring broad angular coverage across the spacecraft's sun-facing plane.}
  \label{fig:tha_fov}
\end{figure}

\begin{figure}[!t]
  \centering
  \includegraphics[width=0.9\textwidth]{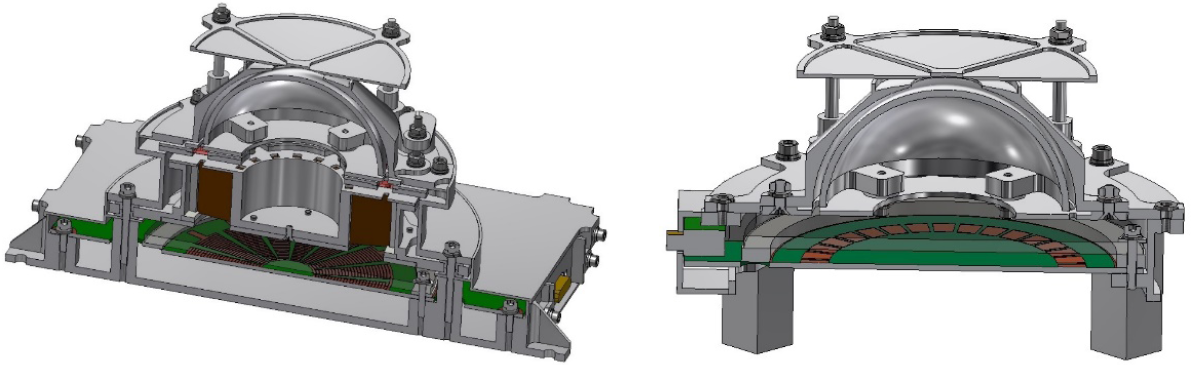}
  \caption{Cutaway view of the Top Hat Analyzer (THA) sensor used in SWIS, showing the hemispherical electrostatic analyzer and microchannel plate (MCP) detectors. The architecture allows energy and angular selection of incident solar wind ions, enabling 3D velocity distribution reconstruction.}
  \label{fig:tha_cutview}
\end{figure}

Aditya-L1 carries seven scientific instruments, including two imaging payloads (Visible Emission Line Coronograph or VELC, Solar Ultra-violet Imaging Telescope or SUIT), two X-ray spectrometers (Solar Low Energy X-ray Spectrometer or SoLEXS, High Energy L1 Orbiting X-ray Spectrometer or HeL1OS), Magnetometer\,(MAG), and two \textit{in-situ} sensors for particle field measurements (Plasma Analyser Package for Aditya\,(PAPA) and \textbf{Aditya Solar Wind Particle Experiment (ASPEX)}) as shown in Fig.\,\ref{fig:tha_fov}. Among these ASPEX is specifically designed to measure solar wind ions across a broad energy spectrum, from 100~eV to 6~MeV/nucleon. ASPEX includes two subsystems: the \textbf{Solar Wind Ion Spectrometer (SWIS)} (AL1-ASPEX-SWIS or SWIS),\cite{prashantswis2024} and the Supra Thermal and Energetic Particle Spectrometer (AL1-ASPEX-STEPS or STEPS)\,\cite{goyal2025steps}. SWIS features dual orthogonally mounted top-hat electrostatic analyzers (THA-1 and THA-2 as in Fig.\,\ref{fig:tha_cutview}) that provide full $360^\circ$ angular coverage in the ecliptic and perpendicular planes, with a narrow out-of-plane acceptance of $\pm1.5^\circ$. It measures ions in the 0.1--20~keV range and a temporal cadence of as high as 5~s, enabling high-resolution observations of solar wind properties. 

During the initial cruise phase of the Aditya-L1 mission, AL1-ASPEX-SWIS was made operational on November 2, 2023. Following its commissioning, it began its observation of solar wind ion energy spectrum. Preliminary datasets revealed clear identification of solar wind streams, validating the instrument’s energy-per-charge\,$({E}/{q})$ resolution and detection thresholds. These early observations confirm the stability and calibration integrity of AL1-ASPEX-SWIS across its dynamic range. Figure \ref{fig:cruise_bulk}, shows the variations in proton number density\,$(n_{\rm d})$, proton bulk speed\,$(v)$, and proton thermal speed $(v_{\rm t})$ in the period between 06--12~November~2023, to demonstrate its nominal performance in-flight as observed before it's routine operations.

Aditya-L1's synergistic role, in coordination with both near-Sun missions and long-term L1 monitors, is vital for bridging existing observational gaps\,\cite{sebastian2025gsics}. Together, these missions pave the way for a deeper understanding of solar wind generation, its propagation through the heliosphere, and its impact on the near-Earth space environment.

The present study highlights the capabilities of the AL1-ASPEX-SWIS instrument aboard Aditya-L1 in characterizing solar wind ion populations with high temporal and directional resolution. As a representative case, we focus on the dynamic interval of 09--15~August~2024, during which an interplanetary coronal mass ejection (ICME) was observed. Leveraging the dual-plane configuration of SWIS—comprising two orthogonally oriented top-hat analyzers, THA-1 and THA-2—we investigate anisotropies in ion energy spectra and temporal variations in alpha particle abundance. This dual-view capability allows SWIS to detect directional features of the solar wind that are often missed in a spin-averaged measurement - such as Wind-SWE-FC.

To assess SWIS’s ability to capture multi-scale kinetic fluctuations, we perform a power spectral density (PSD) analysis of solar wind velocity fluctuations and compare the results with those obtained from Wind-SWE-FC. In addition, we conduct a long-term statistical inter-comparison of key bulk parameters—proton density ($n_{\rm d}$), bulk speed ($v$), and thermal speed ($v_{\rm t}$)—against measurements from Wind-3DP-PLSP, Wind-SWE-FC, and DSCOVR-PlasMag-FC over the 17-month period from January~2024 to May~2025.

The structure of this paper is as follows: Section~2 describes the AL1-ASPEX-SWIS instrument configuration and summarizes the Wind and DSCOVR reference datasets. It also details the fitting methodology used to derive bulk plasma parameters, assuming an isotropic Maxwellian distribution. The subsequent sections present two complementary analyses: an event-based investigation of the August 2024 ICME, and a broader cross-mission comparison assessing the long-term performance and calibration stability of SWIS.

\section{Instrumentation and Data Sources}
Electrostatic analyzers (ESAs) are important instruments in space plasma diagnostics, designed to select charged particles based on their energy-per-charge ratio using electric fields between concentric hemispheres. By sweeping through a range of voltages, ESAs construct detailed ion energy spectra, allowing inference of key plasma properties such as density, velocity, temperature, and composition. AL1-ASPEX-SWIS employs two orthogonally mounted Top-Hat Electrostatic Analyzers (THA-1 and THA-2) fixed on a stable platform as shown in Figure\,\ref{fig:tha_fov}. This dual-plane configuration enables angular coverage across two perpendicular planes, significantly enhancing directional sensitivity to ion flows compared to single-plane instruments. In contrast, Wind-3DP-PLSP utilizes Top-Hat analyzers on a spinning platform to achieve full 3D ion distributions, which is theoretically the optimal configuration for capturing complete angular coverage of incoming particle populations. On the other hand, Wind-SWE-FC and DSCOVR-PlasMag-FC rely on Faraday cups with more limited spatial coverage. These legacy instruments have played a foundational role in shaping our understanding of solar wind structure, composition, and dynamics, and thus serve as valuable benchmarks for validating AL1-ASPEX-SWIS measurements—both during transient events and in long-term solar wind monitoring.

\subsection{Instrument Description}
Each Top-Hat Analyzer (THA) in the AL1-ASPEX-SWIS suite—namely, THA-1 and THA-2—provides full $360^\circ$ azimuthal coverage within its measurement plane, with a narrow acceptance angle of $\pm1.5^\circ$. THA-1 samples particles in the ecliptic plane, while THA-2 observes the orthogonal meridional plane. Both analyzers utilize microchannel plates (MCPs) in a chevron configuration, coupled with position-sensitive anodes to enable simultaneous energy and angular detection. THA-1 is equipped with 16 radial sectors, whereas THA-2 features a 32-sector ring anode, offering enhanced angular resolution.
SWIS covers an energy range of 0.1–20~keV with a nominal resolution of approximately 8\%, enabling discrimination between key ion species such as H$^+$ and He$^{2+}$ in the solar wind. Data are sampled at a cadence of 5~s, aligning closely with the resolution provided by Wind-3DP-PLSP. Unlike Wind, Aditya-L1 is a 3-axis stabilized spacecraft and hence AL1-ASPEX-SWIS-THA1 and AL1-ASPEX-SWIS-THA2 consistently measures in two orthogonal planes. Table\ref{tab:instrument_comparison} summarizes the comparison between the instrument.

\begin{table}[h!]
\centering
\caption{Comparison of Instrument parameters of AL1-ASPEX-SWIS with Wind-3DP-PLSP, Wind-SWE-FC, and DSCOVR-PlasMag-FC Instruments}
\begin{tabular}{|p{2.4cm}|p{3.6cm}|p{3.3cm}|p{2.8cm}|p{2.8cm}|}
\hline
\textbf{Parameter} & \textbf{AL1-ASPEX-SWIS} & \textbf{Wind-3DP-PLSP} & \textbf{Wind-SWE-FC} & \textbf{DSCOVR-PlasMag-FC} \\
\hline
\textbf{Energy Range} & 0.1 – 20 keV & $\sim$0.003 – 30 keV & 0.25 – 8 keV & 0.7 – 8 keV \\
\hline
\textbf{Energy Resolution} & $\sim$8\% & $\sim$20\% & $\sim$5–10\% & $\sim$15–20\% \\
\hline
\textbf{Temporal Cadence} & 5 – 50\,s & 3 – 25\,s & 92\,s & 60\,s \\
\hline
\textbf{Angular Coverage} & 360$^\circ$ in two planes & Full 3D & Limited (single plane) & Single sensor, partial coverage \\
\hline
\textbf{Instrument Geometry} & Two orthogonal Top-Hat Analysers on a stable platform & Top-Hat Analysers on a spinning platform & Faraday cups on a spinning platform & Faraday cup on a stable platform \\
\hline
\textbf{3D Distribution} & No (dual THA orthogonal planes) & Yes & No & No \\
\hline
\textbf{User Configurability} & On-board configurable & No & No & No \\
\hline
\end{tabular}
\label{tab:instrument_comparison}
\end{table}

\subsection{Derivation of solar wind bulk parameters}

Accurate characterization of solar wind plasma requires robust estimation of proton bulk properties $n_{\rm d}$, $v$, and $v_{\rm t}$\,\cite{paschmann1998multi}. In this study, we derive these parameters from differential energy flux measurements by the AL1-ASPEX-SWIS instrument. Assuming that the solar wind proton distribution is approximately Maxwellian and isotropic in the spacecraft frame, we apply two complementary analysis techniques: a parametric Gaussian fit in the velocity space and a non-parametric moment-based estimation. These methods are selected to accommodate varying data quality and observational conditions, offering a flexible framework for deriving consistent plasma moments across a wide range of solar wind regimes.

To derive bulk plasma properties from AL1-ASPEX-SWIS measurements, differential energy flux data are first transformed into velocity space. Assuming isotropic proton distributions, the speed $v$ corresponding to a kinetic energy $E$ is given by $v = \sqrt{2E/m_p}$, where $m_p$ is the proton mass. This transformation facilitates thermal modeling under the assumption of Maxwellian-like velocity distributions. To reduce the influence of instrumental background and low-energy noise, data bins with $v < 280~\mathrm{km~s^{-1}}$ are excluded from the analysis.

As a first step in the parameter estimation process, we compute velocity-space moments of the flux distribution to generate initial guesses for subsequent parametric fitting. Given measured flux values $f_i$ at velocities $v_i$ and bin widths $\Delta v_i$, the proton number density is estimated as $n = \sum_{i=1}^{N} f_i \Delta v_i$, the bulk speed as $v = \frac{1}{n} \sum_{i=1}^{N} f_i v_i \Delta v_i$, and the thermal speed as $\sigma^2 = \frac{1}{n} \sum_{i=1}^{N} f_i (v_i - v)^2 \Delta v_i$, from which the temperature is derived via $T = m_p \sigma^2 / k_B$.

Uncertainties in the derived bulk parameters originate from four principal sources: statistical noise in particle counts, fitting errors from the parametric model, instrument energy resolution, and calibration accuracy. Statistical noise is quantified as $\sigma_{\mathrm{stat}} = \sqrt{N}$, where ${N}$ is the number of detected counts. The uncertainty due to model fitting is extracted from the covariance matrix of the least-squares optimization. Energy resolution contributes an uncertainty estimated as $\sigma_E = 0.08\,{\rm E}$, reflecting the finite bin width and detector response in energy space. Calibration uncertainty is tied to the instrument’s response function ${\rm R}({\rm E}, \theta)$, which varies with both energy and angular incidence. These sources are combined in quadrature to compute the total uncertainty in the differential flux 

To evaluate the quality of empirical fits in scatterplot regressions between SWIS and reference instruments, we use the coefficient of determination, $R^2$, computed via the \texttt{r2\_score} from the \texttt{sklearn.metrics} module. It quantifies the proportion of variance in the observed values explained by the model and is defined as:
\begin{equation}
R^2 = 1 - \frac{\sum_{i=1}^n (y_i - \hat{y}_i)^2}{\sum_{i=1}^n (y_i - \bar{y})^2},
\label{eq:r2_score}
\end{equation}
where $y_i$ are true values, $\hat{y}_i$ are model predictions, and $\bar{y}$ is the mean of $y_i$. An $R^2$ of 1.0 indicates a perfect fit; values $< 0$ imply performance worse than predicting the mean. This metric is used throughout to quantify agreement between SWIS-derived parameters and those from legacy spacecraft.

\section{Results and Discussion}
To evaluate the performance of the AL1-ASPEX-SWIS instrument, this study is organized into six key analytical components. First, we assess SWIS’s operational stability and measurement integrity during its cruise phase prior to L1 insertion. Next, we perform a cross-comparison of solar wind parameters during a dynamic interval (09--15~August~2024), following the commencement of routine science operations after halo orbit insertion and payload verification. This is followed by an investigation into directional differences in solar wind composition, leveraging the dual top-hat analyzers (THA-1 and THA-2) for angularly resolved particle measurements. The fourth section analyzes the temporal variation in helium (Alternatively, alpha or doubly ionized helium or He2+ ion) abundance. We then present a spectral analysis of kinetic energy fluctuations to assess the instrument’s ability to resolve turbulent structures across inertial-range scales. Lastly, a long-term cross-mission comparison is conducted from January~2024 to May~2025 to validate SWIS's consistency and accuracy through regression analyses with Wind and DSCOVR datasets.
\subsection{Operational Readiness During Cruise Phase}

\begin{figure}[!t]
  \centering
  \includegraphics[width=0.8\textwidth]{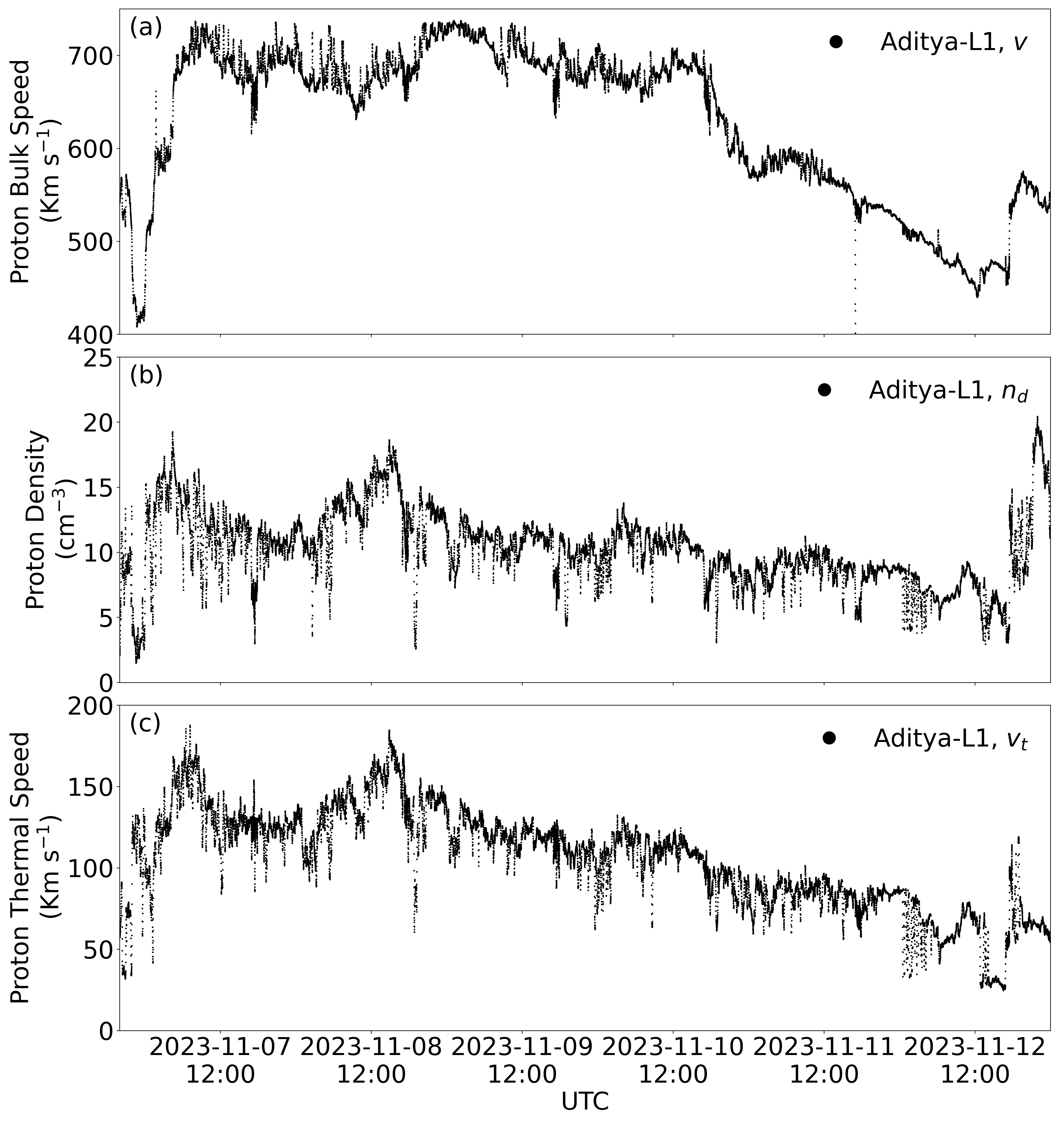}
  \caption{Time series of key solar wind parameters measured by AL1-ASPEX-SWIS during the cruise phase is plotted here, Panels (a-c) shows the variations in proton bulk speed ($v$), number density ($n_{\rm d}$), and thermal speed ($v_{\rm t}$) respectively. These in-situ observations mark some of the earliest solar wind measurements from Aditya-L1 and demonstrate the instrument’s capability to resolve solar wind structures. The measured values lie within expected ranges, validating the nominal performance and calibration integrity in the interplanetary space.}
  \label{fig:cruise_bulk}
\end{figure}

Following its successful launch, Aditya-L1 performed a series of Earth-bound maneuvers before commencing its transfer to the Sun–Earth Lagrange Point~L1. The spacecraft was successfully inserted into a halo orbit around L1 on 6~January~2024. Scientific operations began earlier, on 10~September~2023, with the commissioning of the Suprathermal and Energetic Particle Spectrometer (STEPS), a subsystem of the ASPEX payload. STEPS began collecting energetic particle data near Earth's magnetosphere and bow shock, these observations enabled detailed spectral analysis of energetic ion fluxes from the two STEPS units, providing a unique opportunity to assess the relative roles of external drivers, such as interplanetary coronal mass ejections (ICMEs), and internal geomagnetic processes, such as substorms, in shaping the energetic ion environment within the terrestrial magnetosphere~\cite{chakrabarty2025energetic2mev}.

Aditya-L1 exited Earth's gravitational sphere of influence on 30~September~2023 and subsequently began its cruise toward the Sun–Earth Lagrange Point~L1. During this phase, the AL1-ASPEX-SWIS instrument commenced measurements on 01~November~2023, initiating the recording of ion energy histograms in interplanetary space. These histograms were later processed to derive key solar wind parameters—including bulk speed, thermal speed, and proton density—demonstrating the instrument’s operational readiness during the cruise phase itself. Figure~\ref{fig:cruise_bulk} presents data acquired between 06--12~November~2023, which captures the nominal behavior of the solar wind and offers a representative snapshot of SWIS performance during the cruise phase. This interval followed a period of elevated geomagnetic activity forecast between 04--10~November due to multiple CMEs and a high-speed stream (HSS) as reported by NOAA SWPC (Space Weather Prediction Center), providing an opportunity to assess SWIS under both quiet and transitioning solar wind conditions.

Notably, During this interval a transient structure associated with a halo coronal mass ejection (CME) detected by LASCO on 09~November~2023 at 11:48~UT, exhibited a solar wind speed of approximately 570~km~s$^{-1}$ and a minimum Dst index of $-38$~nT, indicating moderate geo-effectiveness. The AL1-ASPEX-SWIS instrument clearly captured this event during the cruise phase, as shown in Figure~\ref{fig:cruise_bulk}. The data reveal a sharp increase in proton bulk speed, peaking near 600~km~s$^{-1}$ on 12~November, along with an elevated thermal speed approaching 120~km~s$^{-1}$, and proton density reaching approximately 20~cm$^{-3}$. These observations highlight SWIS’s capability to resolve transient solar wind structures with high temporal and energy resolution—demonstrating its scientific readiness prior to routine operations at L1.

Building on the early validation during the cruise phase, we next analyze a complex interval following the halo orbit insertion and in-orbit payload verification, spanning 09--15~August~2024. This period coincided with a significant interplanetary coronal mass ejection (ICME), preceded by a sequence of five Earth-directed CMEs launched between 08--10~August. The resulting impacts on 11--12~August triggered a G4-class (Severe) geomagnetic storm—among the most intense of Solar Cycle~25—marked by disrupted HF communications and elevated satellite drag as reported by NOAA SWPC (Space Weather Prediction Center). These conditions offer a rigorous testbed for evaluating AL1-ASPEX-SWIS performance under enhanced space weather activity.

\subsection{Comparison with other L1 datasets during the Event from 09--15~August~2024}
\begin{figure}[!t]
  \centering
  \includegraphics[width=0.9\textwidth]{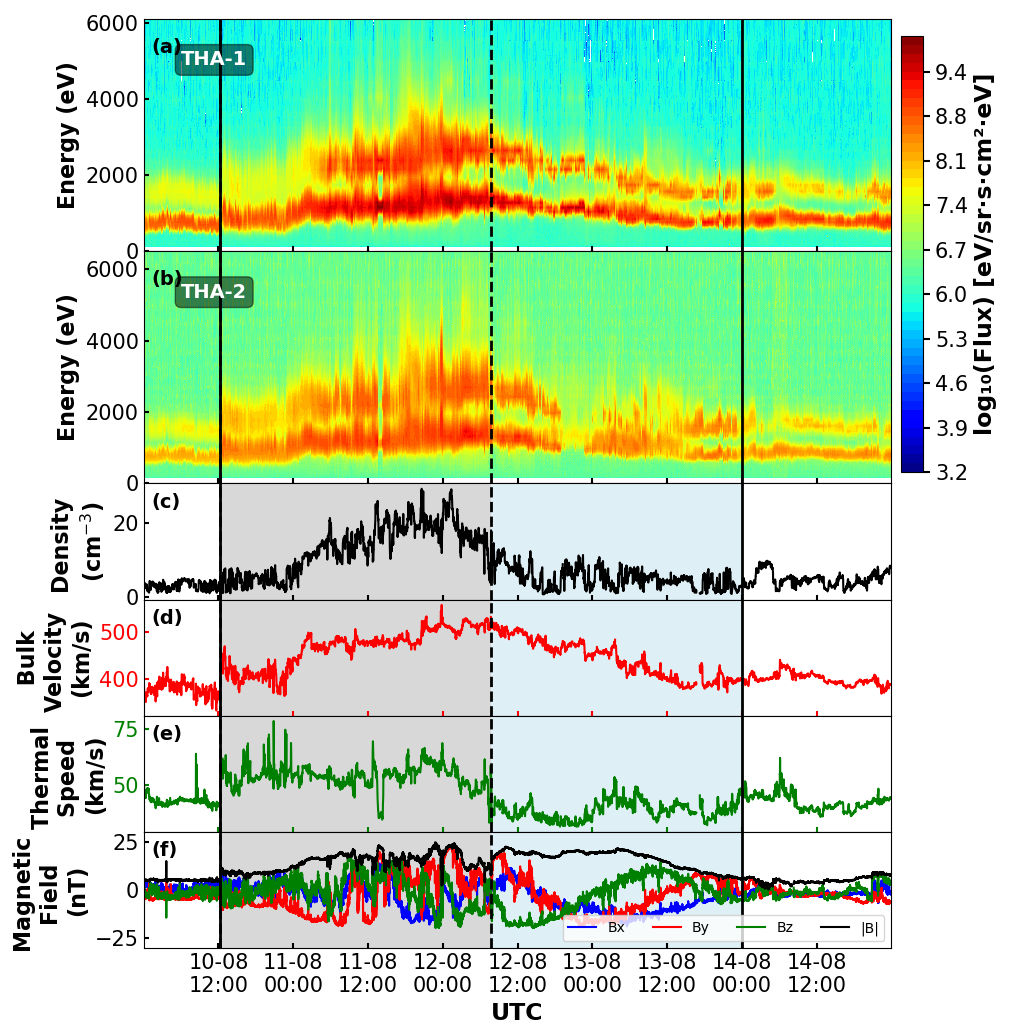}
  \caption{In situ measurements of solar wind parameters observed by AL1-ASPEX-SWIS during 09--15~August~2024. Panels (a) and (b) show the energy histograms from THA-1 and THA-2, respectively. Panel (c) presents the proton number density $n_{\rm d}$ (cm$^{-3}$), while panels (d) and (e) show the bulk speed $v$ and thermal speed $v_{\rm t}$ (both in $\rm km\,s^{-1}$). Panel (f) shows the magnetic field magnitude and its components (Bx, By, Bz) measured by AL1-MAG in GSE coordinates. The solid black line indicates the field magnitude, while blue, red, and green represent Bx, By, and Bz, respectively. Vertical solid black lines mark the ICME boundaries. The first solid black line indicates the shock front, the region between the first solid line and the dashed line represents the sheath, and the region between the dashed line and the final solid black line denotes the magnetic cloud.}
  \label{fig:cme_event}
\end{figure}

This section presents validation of \textit{in-situ} measurements from AL1-ASPEX-SWIS. The observations are cross-validated against concurrent data from Wind-3DP-PLSP, Wind-SWE-FC, and DSCOVR-PlasMag-FC. The analysis demonstrates SWIS’s ability to accurately resolve solar wind parameters and structural transitions during a geoeffective and dynamically evolving solar wind event.

\begin{figure}[!t]
  \centering
  \includegraphics[width=0.8\textwidth]{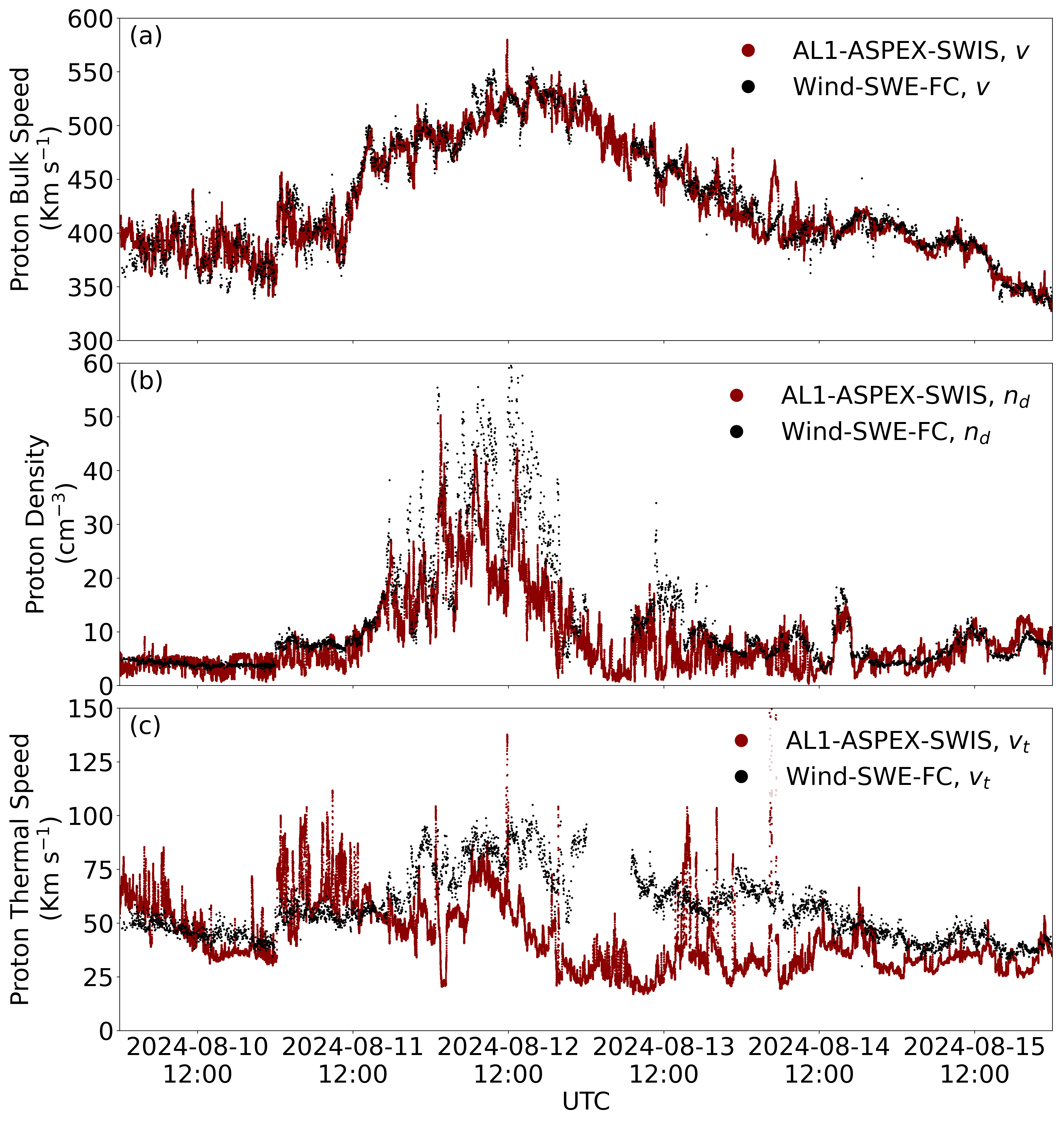}
  \caption{Comparison of proton bulk parameters from AL1-ASPEX-SWIS and Wind-SWE-FC during 09--15~August~2024. The top, middle, and bottom panels display $v$, $n_d$, and $v_t$, respectively. AL1-ASPEX-SWIS data are shown in dark red, and Wind-SWE-FC in black. Excellent agreement is seen in the bulk speed $v$, with consistent trends and timing of key features. The variations in proton density also match well with occasional differences. The variations in thermal speed show significant deviations on some occasions.}
  \label{fig:ts_fc}
\end{figure}

\begin{figure}[!t]
  \centering
  \includegraphics[width=0.8\textwidth]{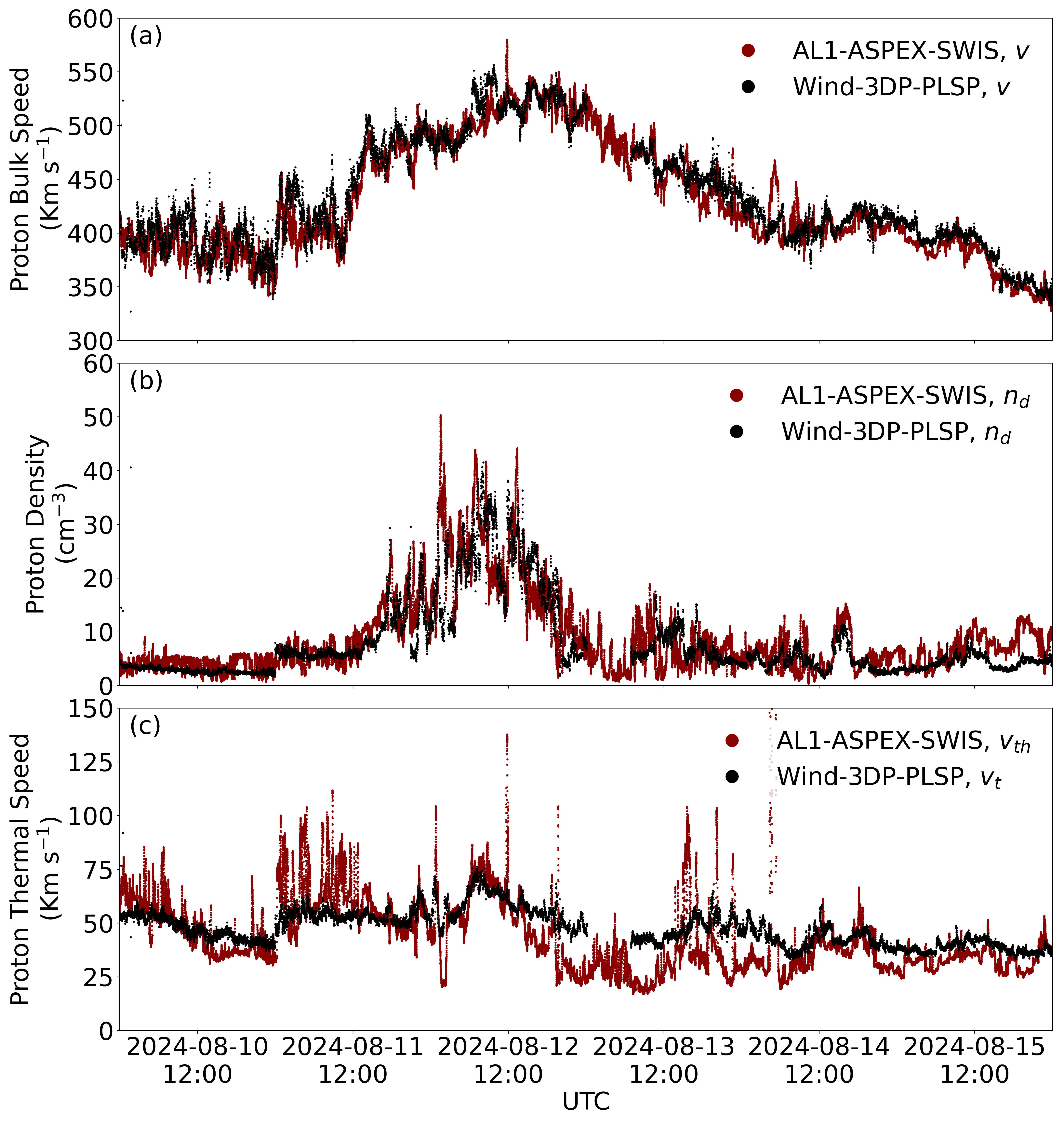}
  \caption{Comparison of solar wind proton parameters from AL1-ASPEX-SWIS and Wind-3DP-PLSP for 09--15~August~2024. From top to bottom, panels show $v$, $n_d$, and $v_t$. AL1-ASPEX-SWIS data appear in dark red, while Wind-3DP-PLSP data are shown in black.}
  \label{fig:ts_plsp}
\end{figure}

\begin{figure}[!t]
  \centering
  \includegraphics[width=0.8\textwidth]{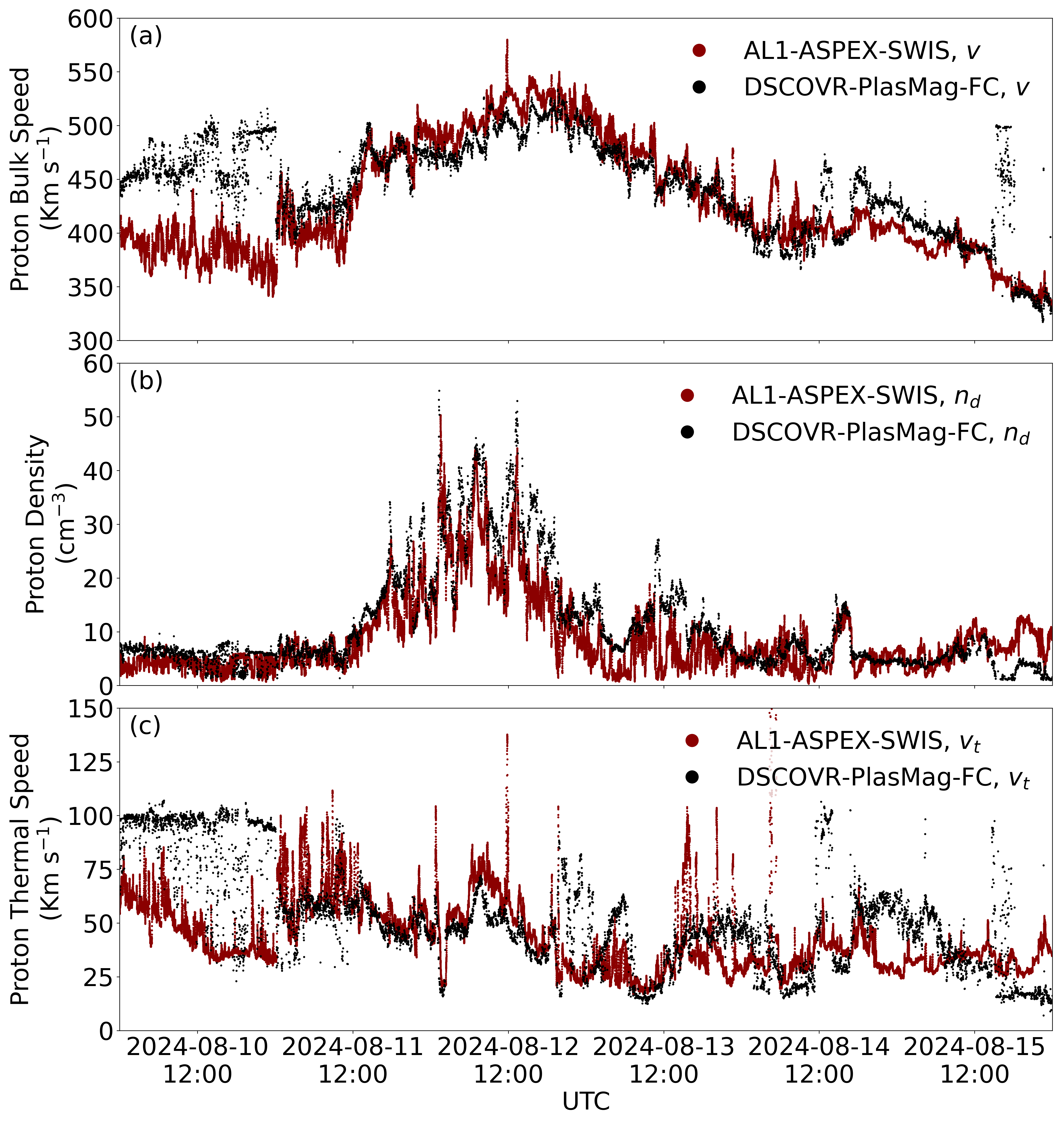}
  \caption{Comparison of proton bulk properties measured by AL1-ASPEX-SWIS and DSCOVR-PlasMag-FC between 09--15~August~2024. Top, middle, and bottom panels show $v$, $n_d$, and $v_t$, respectively. AL1-ASPEX-SWIS data are plotted in dark red; DSCOVR data in black.}
  \label{fig:ts_dscovr}
\end{figure}

Solar transients such as coronal mass ejections (CMEs), solar flares, and high-speed streams are known to drive significant disturbances in the interplanetary medium. Among these, interplanetary CMEs (ICMEs) are the most geoeffective, often causing geomagnetic storms upon interacting with Earth's magnetosphere. To evaluate the response of AL1-ASPEX-SWIS to such a transient event, we show Figure~\ref{fig:cme_event}. This Figure shows energy-flux spectra and key plasma parameters during an ICME that arrived on 11~August~2024. This ICME was associated with a CME observed by LASCO on 7 August 2024 at 03:24 UT and was preceded by multiple high-density CMEs in the days leading up to the event. Clear ICME signatures were detected between 11~August~2024~12:00~UT and 14~August~2024~00:00~UT, including a peak magnetic field of 18~nT and a maximum solar wind speed near 520~km~s$^{-1}$. The average ICME speed was approximately 460~km~s$^{-1}$, and the event reached a minimum Dst of $-188$~nT, indicating strong geo-effectiveness. The energy histograms reveal proton and alpha traces at lower and higher energies respectively. The fluxes are indicated by the color coding. 

Panels (a) and (b) of Figure~\ref{fig:cme_event} show energy-flux histograms from THA-1 and THA-2. Under nominal solar wind conditions, distinct proton and alpha populations are clearly separated. At 09:33~UTC on 10~August~2024, a clear shock arrival is identified by a sudden increase in magnetic field magnitude ($|\vec{B}|$), bulk speed $v$, and thermal speed $v_{\rm t}$—coinciding with enhanced energy flux and visible intermixing of proton and alpha populations, seen as broader and overlapping distributions\,\cite{chakrabarty2022, yogesh2023stream, Kasper_2007, feldman2005sources, alterman2019helium}.

Panels (c), (d), and (e) provide time series of $n_{\rm d}$, $v$, and $v_{\rm t}$, respectively. Panel (f) displays the magnetic field magnitude and components ($B_x$, $B_y$, $B_z$). The sheath region, between the vertical dashed lines, is characterized by turbulent magnetic field fluctuations and sustained enhancements in $v$ and $v_{\rm t}$. The subsequent region—between the dashed line and the final black solid line—corresponds to the magnetic cloud, marked by smooth magnetic field rotation, elevated $|\vec{B}|$, and gradual decreases in $v$ and $v_{\rm t}$, consistent with flux rope expansion. During this phase, $v$ peaks near 550~km~s$^{-1}$, and $v_{\rm t}$ exceeds 80~km~s$^{-1}$. All the $|\vec{B}|$ shows a smooth rotation.

The characteristic features like the elevated and broadened fluxes after the sudden shock feature, overlapping of proton and alpha traces at the sheath due to the enhanced temperatures and clear separations of these traces during the passage of MC are clearly visible. This suggests that all the characteristic features of the encountered ICME are efficiently picked up by AL1-ASPEX-SWIS measurements. This does not mean that every event will be captured by AL1-ASPEX-SWIS (or for that matter, any such instrument) in this conspicuous manner. Since this is a single point measurement in space, the clear signatures of ICME structures will depend on how the spacecraft intercepts the ICME structures. 

To validate the accuracy of SWIS measurements during this interval, we perform a comparison of derived solar wind bulk parameters with those obtained from established L1 monitors. Figures~\ref{fig:ts_fc}--\ref{fig:ts_dscovr} compare AL1-ASPEX-SWIS bulk parameters with measurements from Wind-SWE-FC, Wind-3DP-PLSP, and DSCOVR-PlasMag-FC. Across all datasets, AL1-ASPEX-SWIS shows strong consistency in the bulk speed $v$, particularly aligning well with Wind-SWE-FC. The temporal profile, including the rise to $\sim$550~km~s$^{-1}$ followed by a gradual decline, is clearly reproduced. Trends in $n_d$ and $v_t$ also show strong agreement, although some absolute deviations are evident—likely arising from differences in energy sampling strategies, different locations of the spacecrafts, and data processing methods. Such variations are expected in inter-instrument comparisons and do not undermine the overall qualitative consistency. Notably, a simple exercise (not shown here) suggests that applying a frequency-domain filter further enhances the correlation between the datasets, indicating that high-frequency noise—and potentially spatial decoherence effects due to instrument separation—contribute to the observed discrepancies.

It is crucial to note that comparisons of $n_d$, $v$, and $v_t$ between Wind-SWE-FC and Wind-3DP-PLSP—both aboard the Wind spacecraft—reveal good agreement in $v$, but significant discrepancies in $n_d$ and $v_t$, consistent with earlier findings\,\cite{smith1998wind}. These differences arise despite the instruments being co-located, underscoring the need for caution when using a single dataset as a calibration reference. Rather, these observations highlight the importance of inter-calibration and instrument-aware interpretation in cross-mission studies. Despite these known variances, AL1-ASPEX-SWIS consistently reproduces large-scale solar wind structures, including those associated with the ICME.

In addition to scalar plasma properties, a notable feature of the AL1-ASPEX-SWIS observations is the systematic difference in energy-flux spectra recorded by the two analyzers, THA-1 and THA-2. These differences are not limited to isolated events but appear frequently across various solar wind conditions. This recurring pattern points to the instrument’s sensitivity to directional variations in ion flux. In the following section, we highlight SWIS’s capability to resolve solar wind spatial differences across two orthogonal planes, enabled by its dual-analyzer configuration.

\subsection{Differences in the THA-1 and THA-2 Energy Spectrum: Directional insights}
\begin{figure}[!t]
  \centering
  \hspace*{-0.1\textwidth}\
  \includegraphics[width=1.3\textwidth]{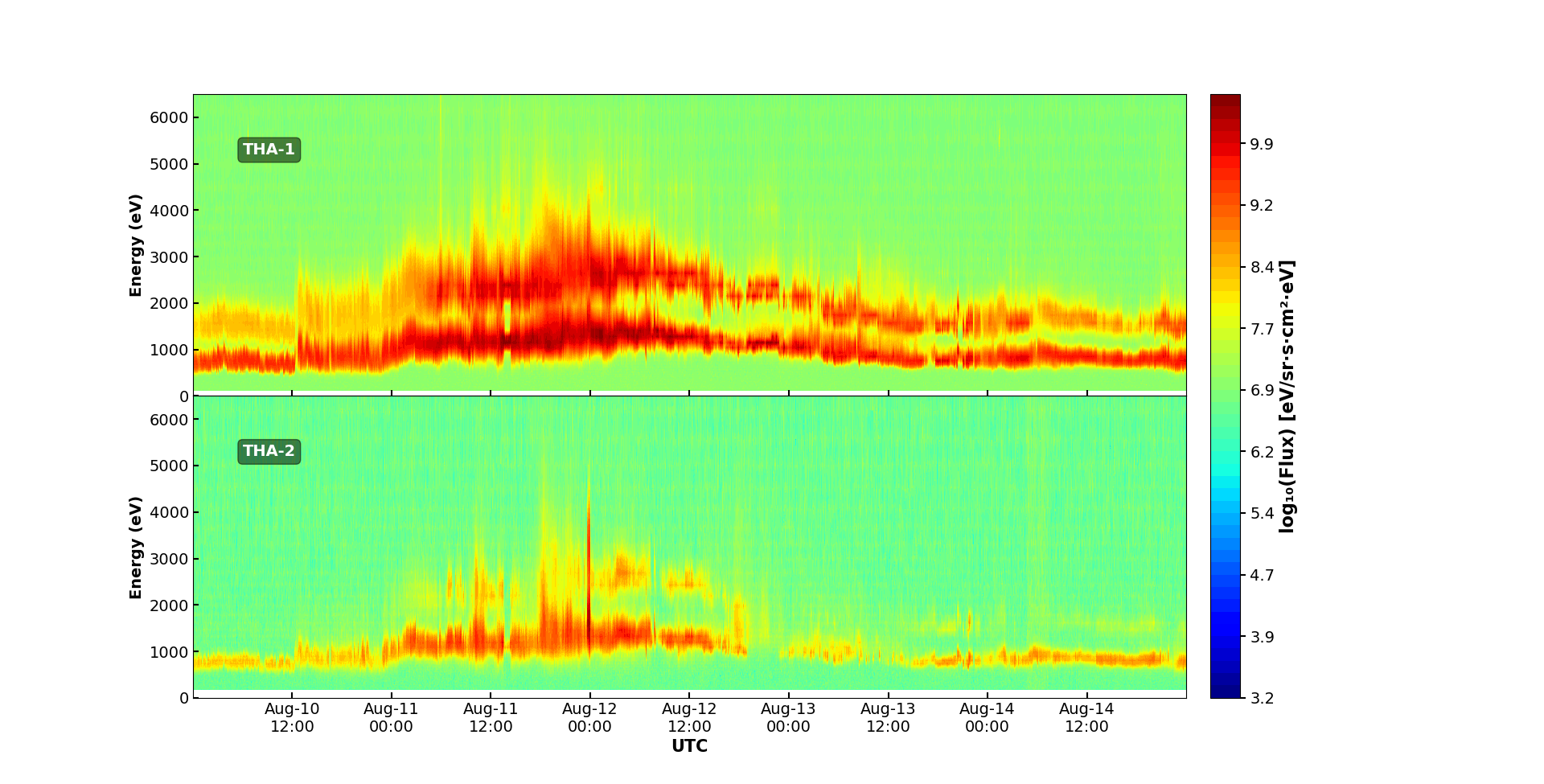}
  \caption{Energy–time spectrograms of differential particle flux measured by AL1–ASPEX–SWIS from THA-1 (top, radial/ecliptic plane) and THA-2 (bottom, perpendicular plane) between 09–15~August~2024. Color indicates $\log_{10}(\text{Flux})$ in units of [eV\,cm$^{-2}$\,sr$^{-1}$\,s$^{-1}$\,eV$^{-1}$]. The observed directional disparity highlights anisotropic ion distributions, particularly during the ICME passage.}
  \label{fig:tha1-tha2}
\end{figure}
The AL1–ASPEX–SWIS instrument aboard Aditya-L1 is specifically designed to address directional anisotropy of solar wind in two mutually orthogonal planes. As discussed earlier, AL1-ASPEX-SWIS consists of two orthogonally oriented electrostatic analyzers—THA-1 and THA-2—that sample ion fluxes in plane. This dual-plane configuration enables simultaneous observations of directional variations in solar wind composition (protons and alphas) and energy distributions, offering a comprehensive view of the spatial structure and evolution of heliospheric plasma at the L1 point.

Figure~\ref{fig:tha1-tha2} displays energy–time spectrograms of ion fluxes measured by THA-1 (top panel, ecliptic plane, $22.5^\circ$ angular resolution) and THA-2 (bottom panel, perpendicular plane, $11.25^\circ$ resolution) from 10–14 August 2024. Measurements are shown for an identical energy range of $\sim$50–6000 eV. The color scale represents $\log_{10}(\text{Flux})$ in units of [eV\,cm$^{-2}$\,sr$^{-1}$\,s$^{-1}$\,eV$^{-1}$]. 

Notably, THA-1 records consistently higher fluxes throughout the observation period, particularly from 10–12 August, which coincides with the passage of an ICME. A distinct bi-modal structure is observed in THA-1 in the 1000–3000 eV range, corresponding to typical alpha particle energies \cite{Kasper_2007, chakrabarty2021, feldman2005sources}. In contrast, THA-2 shows significantly lower fluxes and less distinct ion separation, indicating suppressed alpha signatures. These observations reveal pronounced directional anisotropy in solar wind composition, especially in the alpha component. Although both THA-1 and THA-2 were ground-calibrated and exhibited near-identical laboratory performance \cite{prashantswis2024}, the observed in-flight differences highlight real spatial and directional variability in the solar wind. These aspects hold promising scientific insights which will be taken up in the days to come.

\subsection{Temporal Evolution of the Helium Abundance in the Solar Wind}
\begin{figure}[!t]
  \centering
  \includegraphics[width=0.95\textwidth]{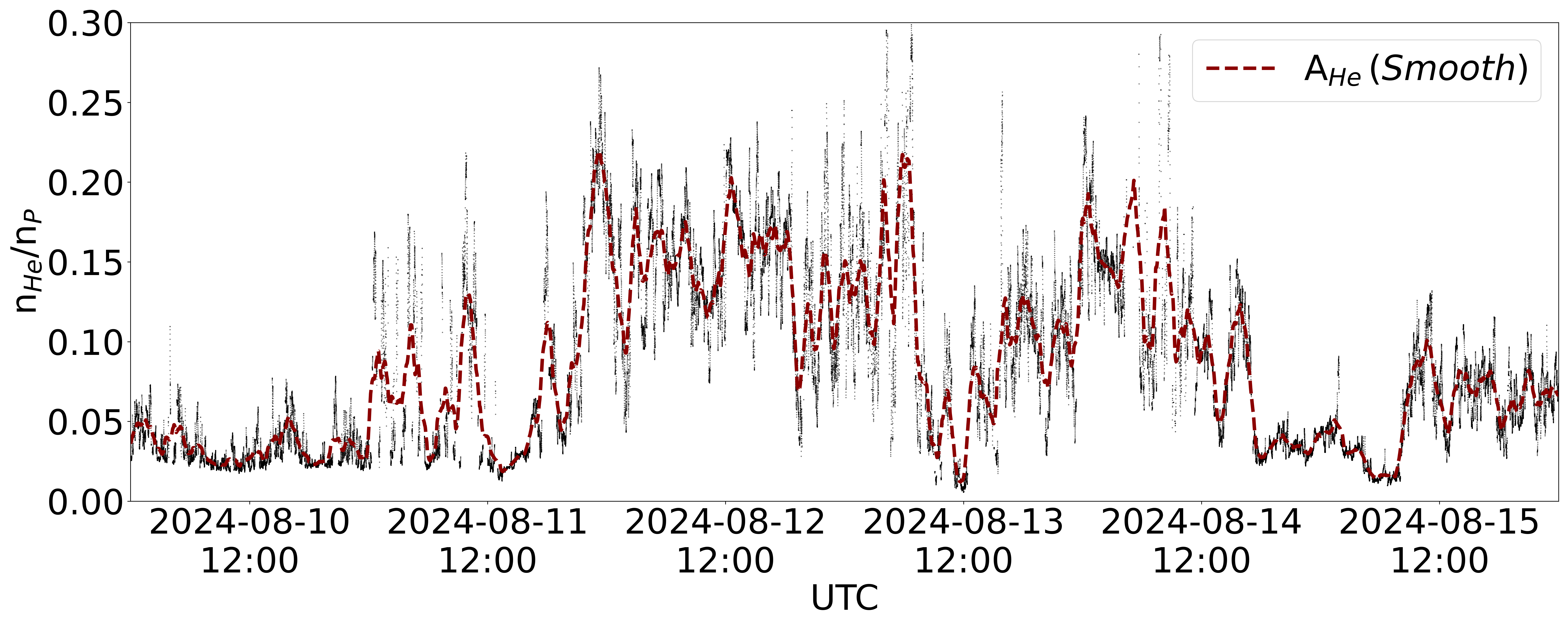}
    \caption{Temporal evolution of the alpha-to-proton density ratio ($n_\mathrm{He}/n_p$) measured by AL1–ASPEX–SWIS from 9 to 15 August 2024. Variations in this ratio reflect changes in solar wind composition and may signify transitions between different solar wind regimes or the passage of transient structures such as interplanetary coronal mass ejections (ICMEs). A smoothed profile (shown in red) using a Savitzky-Golay filter highlights the underlying trends by reducing high-frequency fluctuations.}
  \label{fig:alpha_proton_ratio}
\end{figure}
The variability of helium abundance in the solar wind serves as a key diagnostic of its origin and acceleration processes. Helium ions (He$^{++}$, or alpha particles) typically constitute the second most abundant ion species in the solar wind after protons (H$^+$), contributing about 4--5\% on average, though this value can vary widely. The helium abundance is commonly quantified as
\[
A_{\text{He}} = \left( \frac{n_{\text{He}}}{n_p} \right) \times 100,
\]
where \( n_p \) and \( n_{\text{He}} \) are the number densities of protons and helium ions, respectively.

This ratio is sensitive to the physical conditions in the solar atmosphere, including chromospheric and coronal heating, magnetic topology, and solar wind source regions. In the solar photosphere, helium abundance is about 8.5\% \cite{grevesse1998, asplund2009}, but it is typically depleted in the corona due to First Ionization Potential (FIP) effects \cite{laming2004}. As the solar wind expands into the heliosphere, $A_{\text{He}}$ can range from below 0.1\% to over 10\%, depending on the source region processes, speed, structure, and solar cycle phase \cite{alterman2019helium, chakrabarty2021}. Notably, $A_{\text{He}}$ has been shown to correlate with solar wind speed. A statistical study by \cite{Kasper_2007} introduced the concept of a ``helium vanishing speed'' identifying a threshold near 259~km/s below which helium abundance drops significantly. This behavior suggests that the presence of helium may be linked with the origin of solar wind itself. Enhancements in helium abundance are also characteristic of interplanetary coronal mass ejections (ICMEs), where values of $A_{\text{He}}$ often exceeding 10\% \cite{borrini1982, zurbuchen2006}, indicating compositional fractionation not typical of ambient solar wind. Recent studies \cite{chakrabarty2022} have examined this phenomenon wherein the importance of chromospheric evaporation process and gravitational settling of helium has been highlighted.

Figure~\ref{fig:alpha_proton_ratio} presents the temporal evolution of the alpha-to-proton density ratio ($n_\mathrm{He}/n_p$) as measured by AL1–ASPEX–SWIS between 9 and 15 August 2024. For much of this interval, the ratio fluctuates around a mean value of approximately 0.05, consistent with typical ambient solar wind conditions. However, a marked enhancement is observed beginning late on 11 August and peaking on 12 August, where the ratio exceeds 0.25. This abrupt rise follows a presumed interplanetary shock arrival and is indicative of the passage of a helium-rich plasma structure, likely associated with an ICME.

These variations underscore the sensitivity of AL1–ASPEX–SWIS to subtle changes in composition of the primary species (proton and alpha) in the solar wind and its capability to resolve energy-separated ion populations with high temporal fidelity. By tracking such compositional dynamics, particularly in response to shocks and transients, SWIS can provide crucial link between the solar wind structures measured at L1 and their solar sources.

\subsection{Insights on Kinetic Turbulence}
\begin{figure}[!t]
  \centering
  \includegraphics[width=0.9\textwidth]{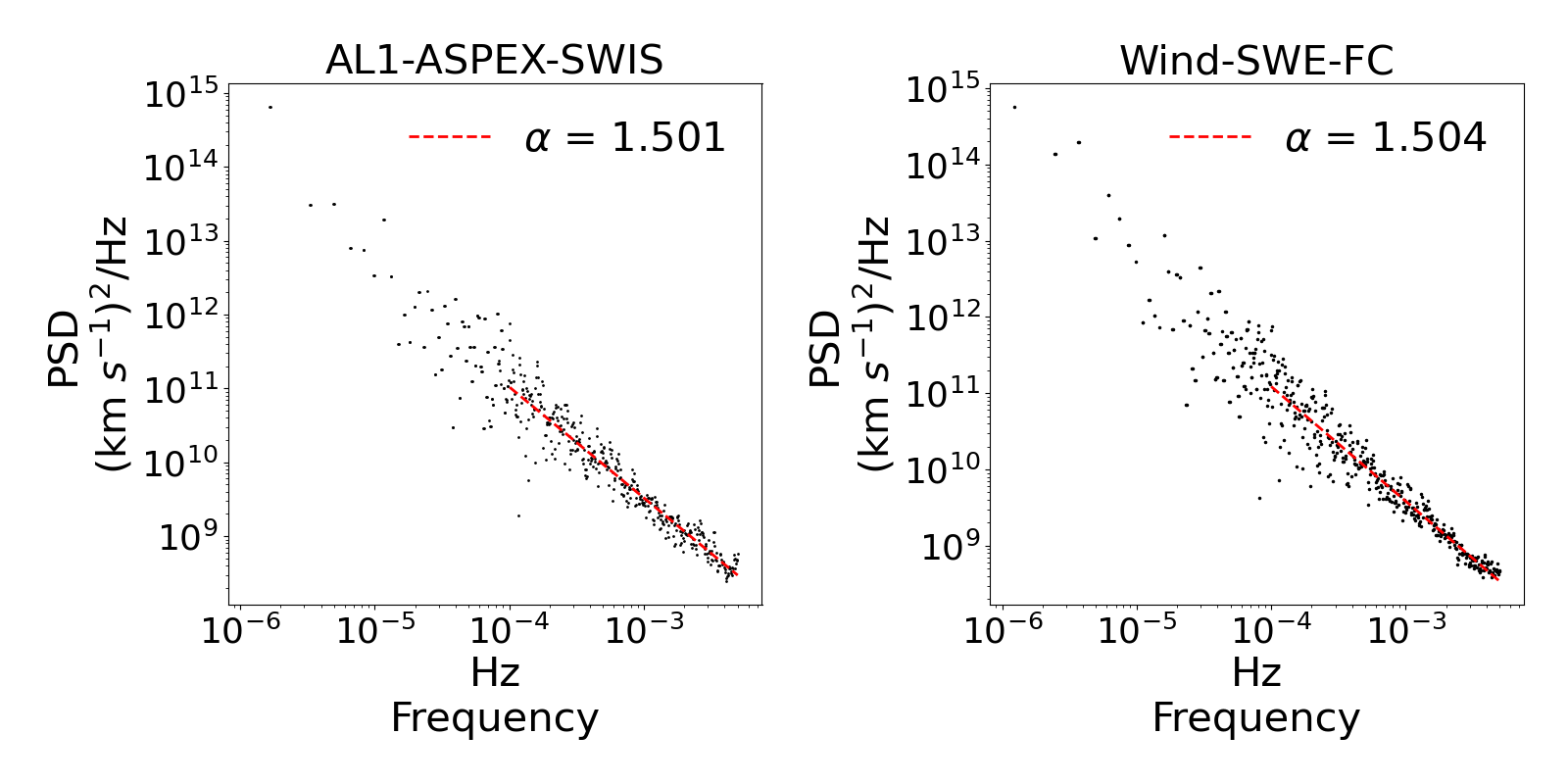}
  \caption{Power spectral densities (PSDs) of solar wind velocity fluctuations from AL1–ASPEX–SWIS (left) and Wind–SWE–FC (right) during 9--15 August 2024. Data were resampled to a 100-second cadence. Power-law fits over the inertial range (\(10^{-4}\)–\(10^{-2}\) Hz) reveal spectral slopes consistent with magnetohydrodynamic turbulence.}
  \label{fig:papoulis_psd_comparison}
\end{figure}
Turbulence plays a central role in mediating the transport and dissipation of energy in the solar wind\,\cite{Podesta_2007}. To assess the ability of AL1–ASPEX–SWIS to capture such turbulent features, we examine power spectral densities (PSDs) of kinetic fluctuations and compare them against established spacecraft observations. Figure~\ref{fig:papoulis_psd_comparison} shows the PSDs of solar wind bulk velocity fluctuations derived from AL1–ASPEX–SWIS data (left panel) and Wind–SWE–FC measurements (right panel) over the period 9--15 August 2024.

The PSDs are plotted on a log--log scale, spanning the frequency range \(10^{-4}\) to \(10^{-2}\)~Hz —corresponding to the inertial range of magnetohydrodynamic (MHD) turbulence. A clear power-law behavior is evident in both datasets. For AL1–ASPEX–SWIS, the inertial-range slope is approximately \(-1.501\), closely matching the Kolmogorov-like MHD turbulence and aligning well with the spectral slope observed in the Wind data.

The strong agreement between the two independent instruments confirms the capability of AL1–ASPEX–SWIS to resolve inertial-range dynamics with high fidelity. These results validate the instrument’s sensitivity, cadence, and stability in capturing fundamental turbulence signatures in the solar wind. This establishes AL1–ASPEX–SWIS as a reliable tool for probing the multi-scale structure of heliospheric plasma and for advancing turbulence studies from its unique vantage at the L1 point.

\subsection{Long-Term Performance Assessment of AL1–ASPEX–SWIS}
\newcommand{\bandimg}[1]{%
  \begin{subfigure}[b]{0.33\textwidth}
    \includegraphics[width=\linewidth]{#1}
  \end{subfigure}\hspace{-0.015\textwidth}
}
\begin{figure}[!t]
  \centering
  \bandimg{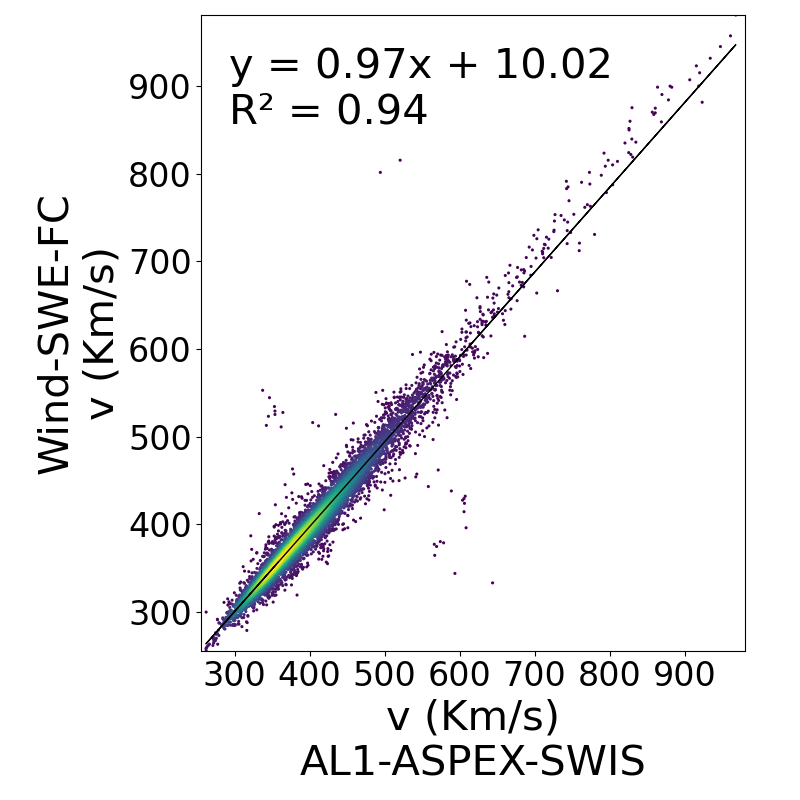}
  \bandimg{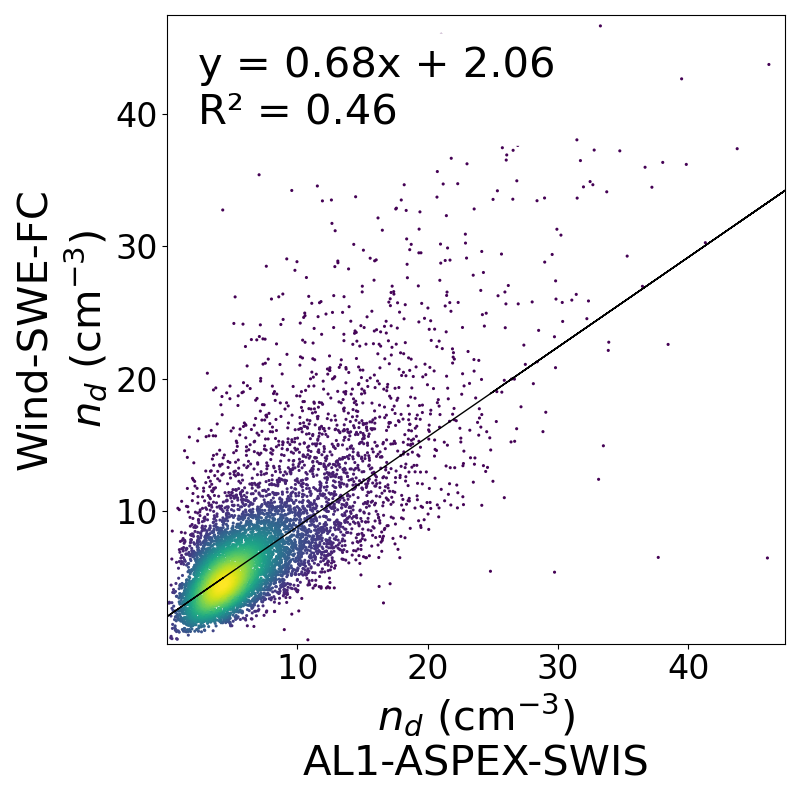}
  \bandimg{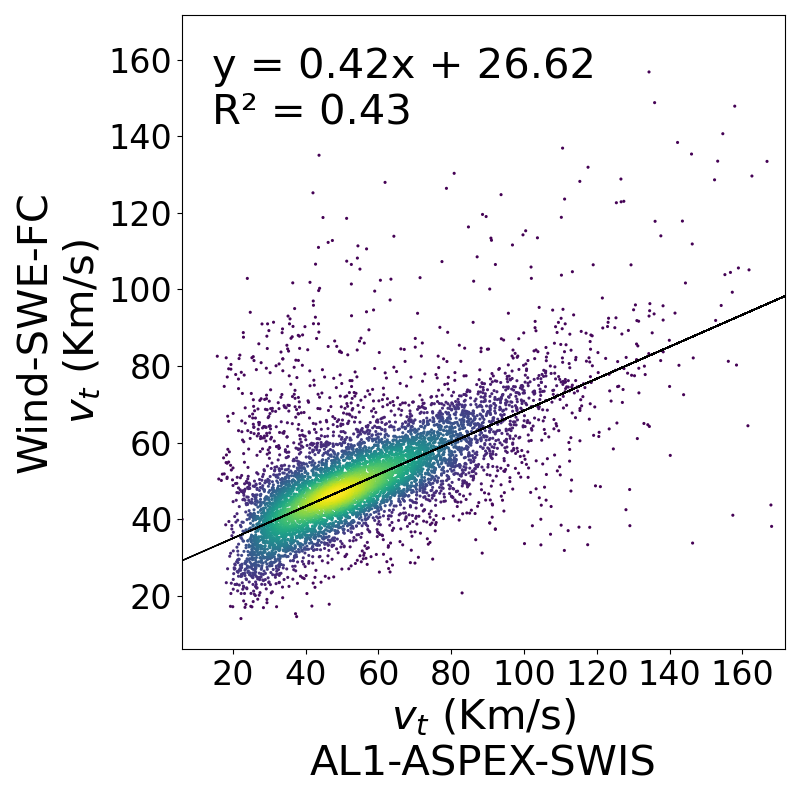}
  \caption{Comparison of $v$ (left), $n_{\rm d}$ (center), and $v_{\rm t}$ (right) between Wind–SWE–FC and AL1–ASPEX–SWIS from January 2024 to May 2025. The $y$-axis shows Wind–SWE–FC values; the $x$-axis shows AL1–ASPEX–SWIS values. Linear regression fits and $R^2$ values are overlaid. Point density is indicated by color, from blue (low) to yellow (high).}
  \label{fig:proton-grid-large}
\end{figure}
To assess the long-term performance and measurement stability of AL1–ASPEX–SWIS, we conducted a cross-mission validation analysis using proton moment data collected between January 2024 and May 2025. During this interval, all relevant housekeeping parameters—including voltages, temperatures, and currents—remained within nominal operational ranges, ensuring consistent instrument behavior and data integrity.

Figure~\ref{fig:proton-grid-large} presents scatter plots comparing key proton bulk parameters measured by AL1–ASPEX–SWIS with concurrent observations from Wind–SWE–FC. Each panel includes a best-fit linear regression line, its equation, and the coefficient of determination ($R^2$), as defined in Eq.~\ref{eq:r2_score}. A color density scale is used to represent the concentration of data points, aiding in visual interpretation.

Among the three compared quantities, bulk speed ($v$) exhibits the strongest agreement with Wind–SWE–FC, yielding an $R^2$ value of 0.94 and a near-unity slope, highlighting the high precision and reliability of AL1–ASPEX–SWIS velocity measurements. Conversely, greater scatter is observed in the density and thermal speed comparisons, particularly against Wind–3DP and DSCOVR–PlasMag. These deviations may arise from inter-instrument calibration differences, differing spatial sampling, or intrinsic spatial inhomogeniety in the solar wind plasma.

Table~\ref{tab:regression-summary} summarizes the regression slopes and $R^2$ values across all instrument pairs and parameters. These results collectively affirm the robust performance of AL1–ASPEX–SWIS for solar wind velocity diagnostics, while also identifying areas—namely, density and thermal speed estimation—where further refinement or cross-calibration could enhance agreement.
\begin{table}[H]
\centering
\caption{Linear regression slopes and $R^2$ values comparing AL1–ASPEX–SWIS proton parameters with Wind and DSCOVR instruments from January 2024 to May 2025.}
\label{tab:regression-summary}
\begin{tabular}{llcc}
\toprule
\textbf{Instrument} & \textbf{Parameter} & \textbf{Slope} & $\mathbf{R^2}$ \\
\midrule
\multirow{3}{*}{Wind–3DP–PLSP}    
& Bulk Velocity   & 1.00 & 0.94 \\
& Thermal Speed   & 0.53 & 0.70 \\
& Proton Density  & 0.38 & -0.35 \\
\midrule
\multirow{3}{*}{Wind–SWE–FC}     
& Bulk Velocity   & 0.97 & 0.94 \\
& Thermal Speed   & 0.42 & 0.43 \\
& Proton Density  & 0.68 & 0.46 \\
\midrule
\multirow{3}{*}{DSCOVR–PlasMag–FC} 
& Bulk Velocity   & 0.82 & 0.44 \\
& Thermal Speed   & 0.70 & 0.07 \\
& Proton Density  & 0.20 & 0.09 \\
\bottomrule
\end{tabular}
\end{table}

\section{Conclusion}
Understanding solar wind properties and characteristics is critical for protecting satellite infrastructure, communication systems, navigation networks, and power grids from the adverse effects of solar activity. Accurate and continuous monitoring of the solar wind at strategic locations like L1 is essential for early warning and detailed characterization of solar transients and ambient plasma conditions. In this context, the AL1--ASPEX--SWIS instrument has been evaluated for its performance in capturing solar wind dynamics through both event-scale diagnostics and long-term monitoring.

Throughout all phases of analysis, SWIS has demonstrated operational stability and scientific integrity, with close agreement to established reference instruments such as Wind and DSCOVR. During the 07~August~2024 ICME, SWIS successfully resolved key solar wind structures, capturing sharp transitions in proton velocity, thermal speed, and number density. These features were temporally aligned with enhancements in energy flux and magnetic field magnitude, affirming SWIS's ability to detect large-scale transients vital for space weather monitoring. Turbulence diagnostics of kinetic-scale fluctuations revealed spectral slopes consistent with magnetohydrodynamic (MHD) turbulence theory, while cross-validation with Wind/SWE--FC confirmed SWIS’s capability in resolving multi-scale plasma dynamics across the inertial range. A 17-month cross-calibration period further affirmed the stability of SWIS measurements. Bulk velocity comparisons yielded excellent agreement ($R^2 \sim 0.94$), with moderate scatter in density and thermal speed attributed to known inter-instrument calibration differences. Additionally, directional flux measurements revealed notable alpha particle anisotropy, with THA-1 consistently observing more structured and elevated fluxes than THA-2. This highlights SWIS’s directional sensitivity and its utility in studying anisotropic particle distributions beyond conventional moment analysis.

In brief, AL1--ASPEX--SWIS provides high-quality, robust solar wind observations with strong temporal resolution and directional fidelity. These capabilities make it a valuable asset for heliophysics research and real-time space weather operations at L1.

\section*{Acknowledgments}

Aditya-L1 is an observatory-class mission which is fully funded and operated by the Indian Space Research Organisation (ISRO). The mission was conceived and realised with the help from various ISRO centres. The science payloads and science-ready data products are realised by the payload PI institutes in close collaboration with ISRO centres. 

We acknowledge the use of data from the Aditya-L1 mission of the Indian Space Research Organisation (ISRO), archived at the Indian Space Science Data Centre (ISSDC). The authors thank the ISRO science and engineering teams involved in the design, development, integration, testing, and operations of the Aditya-L1 mission, and the teams at various ISRO centres whose efforts have made this mission possible.

\section*{Data Availability}
The AL1-ASPEX-SWIS data used in this study are publicly available through the Indian Space Science Data Centre (ISSDC) at \url{https://pradan1.issdc.gov.in/al1/}. The data before the science phase operation are used to generate Figure\,\ref{fig:cruise_bulk} are made available through \url{https://doi.org/10.5281/zenodo.15861770}\,\cite{dataset_2025}. Wind spacecraft data, including SWE\,\cite{WI_H1_SWE} and 3DP instruments\,\cite{WI_PLSP_3DP}, were obtained from NASA's Coordinated Data Analysis Web (CDAWeb) at \url{https://cdaweb.gsfc.nasa.gov}. DSCOVR Faraday Cup Level 2 (1-minute average) data were accessed via NOAA's DSCOVR Data Portal at \url{http://doi.org/10.7289/V51Z42F7}\,\cite{noaa_dscovr_2016}.

\end{document}